\documentclass[prl,aps,showpacs,floatfix,twocolumn,nofootinbib]{revtex4-1}
\usepackage{amsmath}
\usepackage{amsfonts}
\usepackage{amssymb}
\usepackage{revsymb}
\usepackage{graphicx}
\usepackage{mhchem}
\usepackage{siunitx}
\usepackage{bm}
\usepackage{dcolumn}
\usepackage{multirow}

\def\ket#1{|\,#1 \,\rangle}
\def\bra#1{\langle \, #1 \,|}

\def\eval#1{\left\langle #1 \right\rangle}

\begin{document}
\title{Suppression of clock shifts at field-insensitive transitions.}
\author{K. J. Arnold}
\affiliation{Center for Quantum Technologies, 3 Science Drive 2, Singapore, 117543}
\affiliation{Department of Physics, National University of Singapore, 2 Science Drive 3, Singapore, 117551}
\author{M. D. Barrett}
\email{phybmd@nus.edu.sg}
\affiliation{Center for Quantum Technologies, 3 Science Drive 2, Singapore, 117543}
\affiliation{Department of Physics, National University of Singapore, 2 Science Drive 3, Singapore, 117551}
\begin{abstract}
We show that it is possible to significantly reduce quadrupole and tensor polarizability shifts of a clock transition by operating at a judiciously chosen field-insensitive point. In some cases shifts are almost completely eliminated making the transition an effective $J=0$ to $J=0$ candidate.  This significantly improves the feasibility of a recent proposal for clock operation with large ion crystals.  For such multi-ion clocks, geometric constraints and selection rules naturally divide clock operation into two categories based on the orientation of the magnetic field.  We discuss the limitations imposed on each type and how calibrations might be carried out for clock operation.  
\end{abstract}
\maketitle
The realisation of accurate, stable frequency references has enabled important advances in science and technology.  Increasing levels of accuracy and stability continue to be made with atomic clocks based on optical transitions in isolated atoms \cite{AlIon,SrYe,HgIon,SrIon,YbIon,InIon,YbForbidden,RMP,Ludlow}.  However, in the case of single ion clocks, further improvements in accuracy are hindered by their relatively low stability which makes averaging times prohibitively long. 

Recently we have shown that ion clock candidates with a negative differential scalar polarisability, $\Delta \alpha_0$, could operate with large numbers of ions by utilising a magic radio-frequency (RF) trap drive at which micromotion shifts cancel \cite{MDB2}.  Of the candidates reported in the literature that have $\Delta \alpha_0<0$, B$^+$ \cite{Safranova} , Ca$^+$ \cite{Safranova}, Sr$^+$ \cite{Safranova}, Ba$^+$ \cite{Sahoo}, Ra$^+$ \cite{Sahoo}, Er$^{2+}$\cite{Calc1}, Tm$^{3+}$\cite{Calc1}, and Lu$^+$\cite{Calc1}, all but one involve clock states with $J>1/2$.  Methods used to cancel these shifts involve averaging over multiple transitions \cite{MDB1, ZeemanAverage,TransitionAverage}, or multiple field orientations \cite{ItanoQuad}.  In the case of many ions, this leads to inhomogeneous broadening which imposes practical limitations to probe interrogation time and the number of ions one can use \cite{MDB2}.  It is therefore of interest to explore alternative methods to deal with shifts arising from rank 2 tensor interactions. 

Here we focus on those $\Delta\alpha_0<0$ candidates with an upper  $D$ state, specifically Ca$^+$, Sr$^+$, Ba$^+$, Lu$^+$ and Lu$^{2+}$.  We show that shifts from rank 2 tensor interactions can be practically eliminated by operating at a judiciously chosen field-insensitive transition.  In the presence of a magnetic field, states are mixed through the Zeeman interaction which alters the influence of rank 2 tensor interactions relative to the unmixed values.  By example, we illustrate that each candidate has at least one clock transition which becomes field-insensitive and simultaneously the rank 2 perturbations are substantially diminished.  The case of doubly-ionised lutetium has not yet been considered as a viable clock candidate so we also include relevant clock considerations for this ion.

For fixed $J$, the Zeeman interaction is given by
\begin{equation}
H_z=\frac{\mu_B B}{\hbar}\left(g_J J_z+g_I I_z\right),
\end{equation}
with matrix elements
\begin{multline}
\bra{(IJ)F',m_F}H_z\ket{(IJ)F,m_F}=(g_J-g_I)\mu_B B\\
\times(-1)^{J+I+1+m_F}J\sqrt{(2F'+1)(2F+1)}\\
\times\begin{Bmatrix}F & F' & 1\\J & J & I\end{Bmatrix}\begin{pmatrix}J & 1 & J\\-J & 0 & J\end{pmatrix}^{-1}\begin{pmatrix}F & 1 & F'\\-m_F & 0 & m_F\end{pmatrix}\\
+g_I m_F\mu_B B \delta_{F,F'},
\label{Zeeman}
\end{multline}
where we have included $IJ$ in the state notation to specify the order of coupling $I$ and $J$. The Zeeman interaction preserves $m_F$ and eigenstates can be found by diagonalizing the Hamiltonian restricted to a manifold of fixed $m_F$.  Provided perturbations from rank 2 tensor interactions remain small relative to the spacing between energy levels, shifts can be calculated as an expectation value using the new eigenstates.  At low field, a rank 2 perturbation factors into three terms: a state-dependent scalar coefficient depending on the angular momentum quantum numbers, a scalar coefficient depending on the properties of the atom that determines the over-all size of the interaction, and a geometry dependent term that depends only on the strength of the applied fields and their orientation relative to the quantisation axis. Since the Zeeman interaction preserves $m_F$, this separation remains intact provided there is no accidental near degeneracy with neighbouring Zeeman manifolds.  However, the state-dependent coefficient must properly take into account the mixing induced by the Zeeman interaction.  For a given state, the coefficient is determined by the expectation value of the $m_F$ dependent matrix with matrix elements given by
\begin{multline}
H_{F',F}=(-1)^{J+I+m_F}\sqrt{(2F'+1)(2F+1)}\\
\times\begin{Bmatrix}F & F' & 2\\J & J & I\end{Bmatrix}\begin{pmatrix}J & 2 & J\\-J & 0 & J\end{pmatrix}^{-1}\begin{pmatrix}F & 2 & F'\\-m_F & 0 & m_F\end{pmatrix}.
\label{Rank2}
\end{multline}
Matrix elements of the tensor polarisability are given by
\begin{multline}
\bra{F',m_F}H_E\ket{F,m_F}=-\frac{1}{4}H_{F',F} \alpha_{2,J}\\
\times\left\langle3E_z^2-|E|^2\right\rangle
\label{TPshift}
\end{multline}
where $\alpha_{2J}$ is the frequency dependent tensor polarisability for the fine-structure level of interest, and $\eval{\cdot}$ denotes a time average.  Similarly, matrix elements for the quadrupole operator are given by
\begin{multline}
\bra{F',m_F}H_Q\ket{F,m_F}=H_{F',F} \Theta(J)\\
\times  A \left[(3\cos^2\beta-1)-\epsilon \sin^2\beta\left(\cos^2\alpha-\sin^2\alpha\right)\right].
\label{quadshift}
\end{multline}
where $\Theta(J)$ is the quadrupole moment of the fine structure level of interest, $A$ and $\epsilon$ characterize the strength of the applied field gradients, and $\alpha$ and $\beta$ are the Euler angles determining the orientation of the electric field gradient with respect to the quantisation axis.  We note that general matrix elements for the polarizability are given in \cite{PolarizabilityMS} and a slight generalisation of the treatment given in \cite{ItanoQuad} can be used to show they have the same form for the quadrupole interaction.  The matrix elements given in \cite{PolarizabilityCalc} differ from those given in \cite{PolarizabilityMS} due to a different ordering of $I$ and $J$.

\begin{figure}
\begin{center}
\includegraphics[width=3.4in]{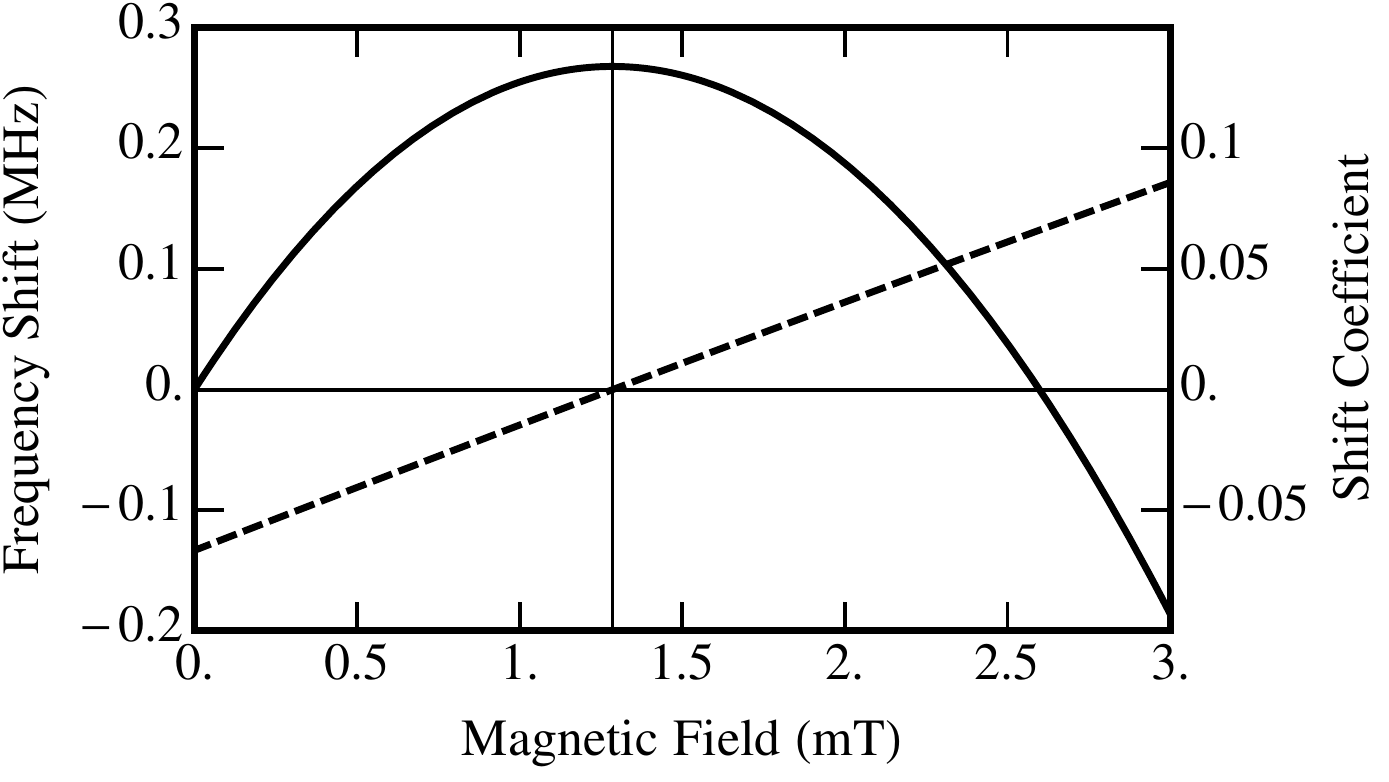}
\caption{The B-field dependence of the $\ket{S_{1/2},4,-3}$ to $\ket{D_{3/2},5,-3}$ transition in $^{43}$Ca$^+$. The solid curve shows the frequency dependence of the transition relative to the zero-field value with the vertical line marking the field-insensitive point of approximately $1.28\,\mathrm{mT}$. The dashed curve shows the field dependence of the corresponding shift coefficient with the horizontal line marking the intercept at the field-insensitive point giving a shift coefficient $C_2\approx -5.8\times 10^{-5}$.}
\label{CaD32}
\end{center}
\end{figure}

For a given atom, it is a simple matter to exhaustively search for field-insensitive points of the clock transition and determine the expectation value of the matrix given in Eq.~\ref{Rank2}.  We refer to this expectation value as the shift coefficient and denote it by $C_2$.  In Fig.~\ref{CaD32}, we illustrate the near coincidence of a field-insensitive point and a zero shift coefficient for $^{43}$Ca$^+$.  Further examples for each candidate are given in table~\ref{table1} where we give the zero-field states associated with the transition, the shift coefficient, and the quadratic Zeeman shift at the field-insensitive point. The table is by no means exhaustive, at least for some of the candidates, and we have listed the most promising transitions for each. 

\begin{table*}
\begin{ruledtabular}
\centering
\begin{tabular}{c c c c c c}
Element & Transition & B (mT) & $\alpha_Z\mathrm{ (kHz/mT^{2})}$ & $C_2$ & Ref. \vspace{2pt}\\
\hline
\vspace{-6pt}\\
$^{43}$Ca$^+$ & $\ket{4,-3}\leftrightarrow\ket{5,-3}$ & 1.28 & -159 & $-5.8\times 10^{-5}$ & \cite{CaHF} \\
$^{43}$Ca$^+$ & $\ket{4,0}\leftrightarrow\ket{3,-2}$ & 0.051 & 58.8 & -0.006 & \cite{CaHF} \\
$^{87}$Sr$^+$ & $\ket{4,-3}\leftrightarrow\ket{3,-1}$ & 5.34 & 304 & -0.004 & \cite{SrSHF,SrDHF} \\
$^{87}$Sr$^+$ & $\ket{4,0}\leftrightarrow\ket{4,0}$ & 12.8 & 69.0 & 0.030 & \cite{SrSHF,SrDHF}\\
$^{137}$Ba$^+$ & $\ket{2,1} \leftrightarrow\ket{2,0}$ & 53.1 & -70.0 & -0.066 & \cite{BaSHF,MDB3}\\
$^{175}$Lu$^{2+}$ & $\ket{4,3}\leftrightarrow\ket{5,3}$ & 22.4 & 9.7 & $-5.6\times 10^{-4}$ & \cite{WJUS}\\
$^{175}$Lu$^{2+}$ & $\ket{4,-3}\leftrightarrow\ket{2,-2}$ & 73.6 & 30.8 & 0.005 & \cite{WJUS} \\
$^{175}$Lu$^+$ & $\ket{7/2,-1/2}\leftrightarrow\ket{9/2,-1/2}$ & 56.6 & 22.1 & -0.035 & \cite{LuHyperfine} \\
$^{176}$Lu$^+$ & $\ket{7,-5}\leftrightarrow\ket{6,-5}$ & $521$ & 29.8 & -0.007 & \cite{LuHyperfine}
\end{tabular}
\caption{Field-insensitive transitions at which the shift coefficient, $C_2$, that scales the rank 2 tensor perturbations is small.  Transitions are identified by their zero-field quantum numbers. All transitions are S$_{1/2}$ to D$_{3/2}$ with the exception of Lu$^+$ which is an $^1$S$_0$ to $^3$D$_2$ transition.  The value of the field at which the transition becomes field insensitive is given in column 3, and the quadratic dependence in given in column 4.  In the final column references from which the hyperfine structure was determined are given.}
\label{table1}
\end{ruledtabular}
\end{table*}

Where possible, analysis is based on experimentally determined hyperfine splittings and this is the case for $^{43}$Ca$^+$, $^{137}$Ba$^+$ and  $^{175}$Lu$^+$.  For $^{87}$Sr$^+$, the experimental value in \cite{SrSHF} for the S$_{1/2}$ level is used but calculations given in \cite{SrDHF} are used for the D$_{3/2}$ level.  Similarly we have relied upon calculations of the hyperfine structure for $^{175}$Lu$^{2+}$ \cite{WJUS}.  For $^{176}$Lu$^+$, we have used experimental values for $^{175}$Lu$^+$ to determine the hyperfine $A$ and $B$ coefficients and rescaled them based on measured nuclear magnetic dipole and electric quadrupole moments.  The results given in table~\ref{table1} have varying levels of sensitivity to calculated values.  However, in general, the Hamiltonian describing Zeeman mixing of the upper state can be scaled by the largest separation of the hyperfine states.  This simply sets a scale for the magnetic field and does not change the form of the eigenenergies and eigenstates as a function of the scaled field.  In as much as the hyperfine splittings are described by the hyperfine $A$ and $B$ coefficients, the rescaled Hamiltonian then depends only on the ratio $B/A$ and this dependence is typically rather weak.  Moreover, for relevant field values, the ground state energies are well described by a comparatively weak quadratic form.  Hence the existence of such coincident points is reasonably robust to small changes in the calculated hyperfine structure.

The case of singly-ionised lutetium is unique in that it is the only candidate that has no hyperfine structure in the ground-state.  The simple structure of the $^3$D$_1$ level does not yield any coincident points and in fact the field-insensitive points typically coincide with extrema in the values of $C_2$.  This is also the case for $^{175}$Lu$^+$ on the $^3$D$_2$ line but it just so happens the extremal value of $C_2$ is also small.   The result given for  $^{176}$Lu$^+$ on the $^3$D$_2$ line involves a much stronger variation in $C_2$ and, consequently has a higher sensitivity to changes in the hyperfine structure compared to others. 

The case of $^{43}$Ca$^+$ offers a unique possibility at low field due to the nuclear spin of $I=7/2$.  In this case the 6J-symbol in Eq.~\ref{Zeeman} vanishes for $F=F'=3$ and the $g$-factor for $F=3$ is given by the nuclear $g$-factor, $g_I$.  It is also the case that the expectation value of any rank 2 tensor vanishes for an angular momentum state $\ket{3,\pm2}$.  Hence, at low fields, these states are inherently free of any significant Zeeman, quadrupole or tensor polarisability shifts.  Furthermore, the quadratic Zeeman shifts of the $\ket{3,\pm2}$ states are anomalously small owing to the small $B$ coefficient for $^{43}$Ca$^+$.   Indeed, the quadratic shift for the transition $\ket{4,0}$ to $\ket{3,-2}$ is almost entirely due to the ground-state. With the very small linear Zeeman shift, the transition becomes field-insensitive at approximately $50\,\mathrm{\mu T}$.  This field gives a small amount of Zeeman mixing which accounts for the shift coefficient of $-0.006$ given in the table.

The result for the $\ket{4,-3}\leftrightarrow\ket{5,-3}$ transition in $^{43}$Ca$^+$ is also a consequence of the $I=7/2$ nuclear spin.  In this case the difference in $g-$factors for the two states is just $0.01$. This provides a field-insensitive point at low field which can be treated within the framework of perturbation theory.  Neglecting the quadratic dependence of the ground-state, the magnetic field value at the field-insensitive point is linear in the hyperfine separation of the  $F=4,$ and $5$ upper levels.  To the same approximation, the mixing of the two upper levels at the field-insensitive point is independent of the hyperfine separation and the zero-field shift coefficient of $C_2=-1/15$ is shifted to $\approx -0.0083$.  Including the quadratic dependence of the ground state decreases the quadratic dependence of the transition and increases the $B$-field of the field-insensitive point.  The associated increase in Zeeman mixing brings $C_2$ to approximately zero.  The same calculation applies to the corresponding transition in $^{175}$Lu$^{2+}$ with a sign change in $m$ corresponding to a sign change of the nuclear magnetic moment.  Similar reasoning also applies to the $\ket{4,-3}\leftrightarrow\ket{3,-1}$ transition in $^{87}$Sr$^+$ although this transition involves stronger mixing of the upper states.  Consequently, for all three of these transitions, the $C_2$ coefficient is insensitive to exact values of the hyperfine splittings.

In the sequence of alkaline-earth ions, the differential scalar polarisability, $\Delta \alpha_0$, becomes increasingly negative with increasing atomic mass.  Doubly-ionized lutetium takes the role of a heavy alkaline-earth-like ion but the extra ion charge offsets this trend.  From published matrix elements \cite{LuIIICalc}, we estimate $\Delta \alpha_0\approx -20.7$ with a tensor contribution of $\alpha_{2,J}=-5.2$ where both results are given in atomic units.  As the scalar polarizabilities of ground and excited states are significantly different (28.0 and 7.3 respectively), we expect the estimate of $\Delta\alpha_0$ to be reasonable.  Using the flexible atomic code \cite{FAC}, we have calculated the lifetime of the D$_{3/2}$ level to be approximately 160 seconds.  The long lifetime, together with the near vanishing shift coefficient of the $\ket{4,-3}\leftrightarrow\ket{5,-3}$ transition, makes this a particularly interesting candidate for a multi-ion clock.

In the multi-ion clock proposal \cite{MDB2}, clock interrogation along the RF null axis of a linear Paul trap is required to avoid micromotion-induced depletion of the probe coupling.  For the E2 transitions considered here, selection rules constrain the orientation of the magnetic field that one can use.  This leads to two types of operation: one in which the field is aligned to the trap axis, and the other in which it rotated by a non-zero angle. The former only applies to those transitions with $|\Delta m|=1$ but in either case, the ultimate performance will depend on the alignment of the magnetic field with respect to the trap axis.

With the Euler angles $\alpha$ and $\beta$ as defined in Eq.~\ref{quadshift}, the tensor polarisability shift due to an RF electric field $\mathbf{E}=(E_x,E_y,0)$ is given by
\begin{multline}
\frac{\delta \nu}{\nu} = -\frac{C_2 \alpha_{2,J}}{4 h \nu} \Bigg(-\frac{1}{4}\left(3\cos^2\beta-1\right)|\mathbf{E}|^2\\
+\frac{3}{4}\sin^2\beta\left(\cos2\alpha \left(E_x^2-E_y^2\right)-2\sin2\alpha E_x E_y\right)\Bigg),
\label{TPshift2}
\end{multline}   
where we have assumed the RF field to be purely transverse to the trap axis as considered in \cite{MDB2}.  In a Coulomb crystal, the second term in Eq.~\ref{TPshift2} has a mean value of zero, provided $\eval{E_x^2}=\eval{E_y^2}$, and gives a number-dependent inhomogeneous broadening of the line.  In a spherically symmetric crystal the broadening would be symmetric and would not give rise to additional shifts associated with changes to the line shape.  Small differences in $\eval{E_x^2}$ and $\eval{E_y^2}$ could be tolerated by setting $\alpha=\pi/4$ which means the projection of the magnetic field on the $xy$-plane is at $45^\circ$ to both the $x$ and $y$ axis.  The first term in Eq.~\ref{TPshift2} has a mean value proportional to $N^{2/3}$. This gives rise to a number-dependent shift of the transition.

When the field is aligned to the trap axis, the number-dependent broadening can be compensated by a slight adjustment of the magic RF frequency \cite{MDB2}.  The only other broadening mechanism is from the crystal-induced quadrupole shifts which are independent of number and heavily reduced by $C_2$.  In practice one would simply increase the number ions, tuning the RF frequency to eliminate variations in the clock frequency with number, and tuning the field alignment to null any observable number-dependent broadening. Hence, this operation allows for arbitrary levels of stability with only a residual quadrupole shift induced by DC electric field gradients.  Since the magnetic field would be calibrated to the trap axis, the dominant contribution from the DC confinement could be calibrated by measuring the trap frequencies leaving only a contribution from stray fields as these need not be aligned with the trap axis.

Off-axis operation is needed when $|\Delta m|=0,2$.  Assuming the magic RF frequency has been pre-calibrated by another method, the orientation of the magnetic field can then be tuned so that $3\cos^2\beta-1=0$ by eliminating any observable number-dependent shifts in the clock frequency.  From Eq.~\ref{quadshift}, the magnetic field would also be oriented such that the residual quadrupole shift arising from the DC confinement fields is also canceled.  Errors in the value of the magic RF frequency will lead to errors in field alignment and a shift of the clock frequency from the DC confinement field.  Further quadrupole shifts from stray DC fields will arise as before.  However, it can be expected that these would be much less than the applied fields. 

For off-axis operation, number-dependent broadening limits the achievable stability.  The broadening scales quadratically with the size of the crystal, which scales as $N^{1/3}$.  The maximum interrogation time then scales as $N^{-2/3}$ leading to an instability with a weak $N^{-1/6}$ scaling.  For all practical purposes this can be considered constant once the broadening begins to degrade the Ramsey fringe or the line shape.  For the spherically symmetric case considered in \cite{MDB2}, the broadening is characterised by 
\begin{equation}
\Delta f = \frac{C_2}{4}\left|\frac{\alpha_{2,J}}{\Delta \alpha_0}\right|\left(\frac{Z^2 \alpha \hbar \omega_z }{m c^2}\right)^{2/3} f_c N^{2/3},
\label{df}
\end{equation}
where  $\alpha$ is the fine structure constant, $Z$ is the charge number of the ion, $\omega_z$ is the trapping frequency along the axial direction, and $f_c$ the clock frequency. Based on simulations, the fringe contrast is reduced to $\sim 80\%$ for a Ramsey time of $1/\Delta f$.  Thus Eq.~\ref{df} provides an effective bound on the number of ions for a given interrogation time.  The dependence on $\Delta \alpha_0$ arises because it determines the magic RF frequency which, in turn, determines the strength of the electric field needed for a particular confinement.  In terms of broadening, this favours candidates in which the ratio $\alpha_{2,J}/\Delta \alpha_0$ is small.  For  $^{43}$Ca$^+$, and $^{175}$Lu$^{2+}$ these considerations are largely irrelevant with the very small $C_2$ value providing an effective $J=0$ to $J=0$ transition.

In conclusion, we have shown that field-insensitive points of clock transitions may be found at which perturbations from rank 2 tensor interactions are significantly diminished.  In the case of $^{175}$Lu$^{2+}$, this approach provides a clock candidate that is essentially field-free except for a $9\,\mathrm{kHz/mT^2}$ quadratic Zeeman shift.  Although this is large relative to other clock candidates it must be remembered that clock operation would be at the field-insensitive point so that only rms field variations need to be accounted for.  This approach significantly improves the feasibility of a recent proposal for clock operation with large ion crystals.

We would like to thank W. R. Johnson and U. I. Safranova for providing us with calculations of the hyperfine structure constants for \ce{^{175}Lu^{2+}}. This research is supported by the National Research Foundation, Prime MinisterÕs Office, Singapore and the Ministry of Education, Singapore under the Research Centres of Excellence programme.  It is also supported in part by A*STAR SERC 2015 Public Sector Research Funding (PSF) Grant (SERC Project No: 1521200080).
\bibliographystyle{apsrev4-1}
\bibliography{ShiftSuppression}
\end{document}